\documentclass[prl,aps,twocolumn,showpacs]{revtex4}

\usepackage{amsmath}
\usepackage{bm}
\usepackage{graphicx}

\begin{document}

\title{Von K\'arm\'an vortex street in a Bose-Einstein condensate}

\author{Kazuki Sasaki}
\author{Naoya Suzuki}
\author{Hiroki Saito}
\affiliation{
Department of Applied Physics and Chemistry, University of
Electro-Communications, Tokyo 182-8585, Japan
}

\date{\today}

\begin{abstract}
Vortex shedding from an obstacle potential moving in a Bose-Einstein
condensate is investigated.
Long-lived alternately aligned vortex pairs are found to form in the
wake, as for the von K\'arm\'an vortex street in classical viscous
fluids.
Various patterns of vortex shedding are systematically studied and the
drag force on the obstacle is calculated.
It is shown that the phenomenon can be observed in a trapped system.
\end{abstract}


\maketitle

The formation of a train of alternate vortices in the wake past an
obstacle, known as the von K\'arm\'an vortex street, is a ubiquitous and
intriguing phenomenon in fluids.
Since the pioneering experimental study by B\'enard~\cite{Benard} and
theoretical consideration by von K\'arm\'an~\cite{Karman}, numerous
studies have been made on the phenomena of vortex street
formation~\cite{Williamson}.

The behavior of a viscous fluid flowing past an obstacle is determined by
the Reynolds number Re, which is a dimensionless parameter that includes
the kinematic viscosity.
The wake of a cylinder is steady for Re below around 50, and a vortex
street emerges for $10^2 \lesssim {\rm Re} \lesssim 10^5$, which becomes
turbulence for a larger Re.
This implies that the viscosity plays an important role in vortex street
formation in classical fluids.
For superfluids, however, the Reynolds number cannot be defined because
of the absence of viscosity.
Moreover, the vortex quantization makes superfluid dynamics quite
different from classical fluid dynamics.
Therefore, it is not obvious whether instability of the wake and
subsequent vortex street generation occur in superfluids.
According to von K\'arm\'an's theory~\cite{Karman,Lamb}, a vortex street
is expected to be very long-lived in inviscid fluids, once it is
created.

In this Letter, by numerically solving the Gross-Pitaevskii (GP)
equation, we show that long-lived alternately aligned vortex pairs are
formed in the wake of an obstacle potential moving in a Bose-Einstein
condensate (BEC), which is similar to the von K\'arm\'an vortex street
in classical fluids.
Mean-field analysis of systems of a BEC with a moving potential has been
performed by many authors from the viewpoints of drag
force~\cite{Frisch,Wini}, vortex dynamics near the 
cylinder~\cite{Nore00}, critical velocity~\cite{Sties}, supersonic
flows~\cite{El,Carusotto}, and multicomponent
systems~\cite{Susanto,Rodrigues}.
However, vortex street formation was not found in these studies,
probably because the parameter region for which a vortex street emerges
is narrow, as shown later.

We consider a BEC of atoms with mass $m$ and an obstacle potential $V$
moving in the $-x$ direction at a velocity $v$.
In the mean-field theory, the condensate is described by the macroscopic
wave function $\psi$ obeying the GP equation given by
\begin{equation} \label{GP}
i \hbar \frac{\partial \psi}{\partial t} = -\frac{\hbar^2}{2m} \nabla^2
\psi + V \psi + g |\psi|^2 \psi,
\end{equation}
where $g = 4 \pi \hbar^2 a / m$ with $a$ being the $s$-wave scattering
length of the atoms.
We employ a Gaussian potential with peak strength $V_0$ and radius $d$
moving in the $-x$ direction at a velocity $v$ as $V = V_0 \exp\{-[(x + v
t)^2 + y^2] / d^2 \}$.
Normalizing space and time by $\hbar / (m g n_0)^{1/2}$ and $\hbar / (g
n_0)$, where $n_0$ is the atom density far from the potential, we can
eliminate the interaction parameter $g$ from Eq.~(\ref{GP}).
We numerically solve Eq.~(\ref{GP}) in two dimensions under the periodic
boundary condition using the pseudo-spectral method.
The initial state is the stationary state of Eq.~(\ref{GP}) with $v = 0$
plus a small amount of noise to break the symmetry.

\begin{figure*}[t]
\includegraphics[width=16cm]{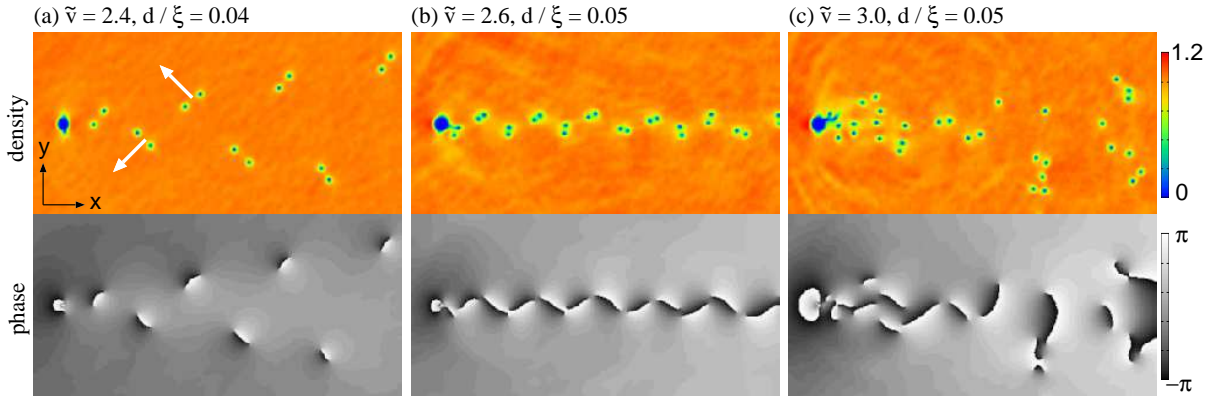}
\caption{
(Color) Density and phase distributions of a condensate past an obstacle
potential.
The velocity and potential width are $(\tilde v, d / \xi) = (2.4, 0.04)$
in (a), $(2.6, 0.05)$ in (b), and $(3.0, 0.05)$ in (c), where $\tilde v
= v [10^3 m / (g n_0)]^{1/2} / (2 \pi)$ and $\xi = \hbar [10^3 / (m g
n_0)]^{1/2}$.
The white arrows in (a) indicate the directions in which the
vortex-antivortex pairs move.
The density is normalized by $n_0$.
The field of view is $6 \xi \times 3 \xi$.
In the numerical calculation, a $32 \xi \times 8 \xi$ space is
discretized into $4096 \times 1024$.
}
\label{f:fig1}
\end{figure*}
Figure~\ref{f:fig1} shows typical wakes flowing past an obstacle
potential with $V_0 / (g n_0) = 100$.
For a sufficiently small velocity $v$, the flow around the obstacle is
a steady laminar flow and no quantized vortex is created.
When the velocity $v$ exceeds a critical velocity, which depends on the
interaction strength and the shape of the potential, vortex-antivortex
pairs are created~\cite{Frisch}.
The critical velocity for the vortex creation is of the order of the
speed of sound $(g n_0 / m)^{1/2}$.
When a created vortex-antivortex pair separates from the potential, the
flow velocity around the potential again exceeds the critical velocity
and a subsequent vortex-antivortex pair is created.
A train of vortex-antivortex pairs is thus generated behind the
potential.
Since a symmetric double row of vortices is unstable~\cite{Lamb}, the
vortex pairs are dislocated sinuously as shown in Fig.~\ref{f:fig1} (a).
Nore {\it et al.}~\cite{Nore93} showed that such a staggered vortex
pattern is formed if a double row of vortices is prepared with an
appropriate perturbation.
Since a pair of point vortices with circulations $\pm h / m$
($h$: Planck's constant) and distance $d$ moves in the direction
perpendicular to a line between the pair at a velocity $\hbar / (m
d)$~\cite{Lamb}, the alternately inclined vortex pairs move in two
directions [white arrows in Fig.~\ref{f:fig1} (a)], forming a V-shaped
wake as in Fig.~\ref{f:fig1} (a).
The divergence of the wake is significant for large $v$, which forms a
pattern similar to supersonic flow~\cite{El,Carusotto}.

Figure~\ref{f:fig1} (b) shows the main result of this study.
The significant difference from Fig.~\ref{f:fig1} (a) is that the
vortices in a pair created by the obstacle potential at a time have the
same circulation.
Since two point vortices having the same circulation $h / m$ rotate
around their center at an angular frequency $2 \hbar / (m d^2)$ without
changing their distance~\cite{Lamb}, the created vortex pairs in
Fig.~\ref{f:fig1} (b) remain bound and rotate.
The pairs with opposite circulations are alternately released from the
obstacle potential to form a train of vortex pairs, resembling a von
K\'arm\'an vortex street.
In contrast to the vortex arrangement originally considered by von
K\'arm\'an, in which isolated point vortices are aligned, the vortex
pairs constitute the vortex street in the present case.
We find from Fig.~\ref{f:fig1} (b) that the distance between the two
vortex rows is $b \simeq 0.24 \xi$ and the distance between two pairs
in a row is $\ell \simeq 0.87 \xi$ on average, and hence $b / \ell
\simeq 0.28$, where $\xi = \hbar [10^3 / (m g n_0)]^{1/2}$.
This ratio is in good agreement with the stability condition of von
K\'arm\'an's vortex arrangement $b / \ell = \pi^{-1} \cosh^{-1} \sqrt{2} 
\simeq 0.28$~\cite{Karman,Lamb}.
In fact, the vortex street in Fig.~\ref{f:fig1} (b) survives at least $t
> 10^3 \hbar / (g n_0)$.
The vortex street in Fig.~\ref{f:fig1} (b) moves in the $-x$ direction
at a velocity $\simeq 0.14 (g n_0 / m)^{1/2} \simeq 0.8 h / (\sqrt{2}
\ell m)$.
The velocity of von K\'arm\'an's point vortices, in which each vortex
has a circulation $2 h / m$, is $h / (\sqrt{2} \ell
m)$~\cite{Karman,Lamb}.
For large $d$ and $v$, the periodicity in the wake seems to disappear
[Fig.~\ref{f:fig1} (c)].
We have numerically confirmed that similar wakes are also obtained for a
disk-shaped potential ($V = \infty$ for $r < d$ and $V = 0$ for $r >
d$).

\begin{figure}[t]
\includegraphics[width=8cm]{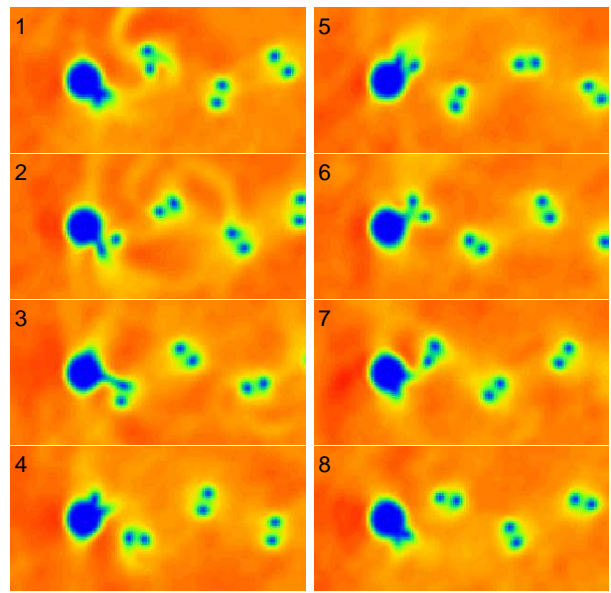}
\caption{
(Color) Serial snapshots of the density profiles for the parameters in
Fig.~\ref{f:fig1} (b) in the frame moving with the potential.
The time interval is $10 \hbar / (g n_0)$.
The field of view is $2 \xi \times \xi$.
The color scale is the same as that in Fig.~\ref{f:fig1}.
}
\label{f:fig2}
\end{figure}
Figure~\ref{f:fig2} shows the dynamics of vortex street formation just
behind the obstacle potential.
The pairs of vortices are released obliquely backward left and right
with alternate circulations.
This vortex street formation behavior is quite different from that in
classical viscous fluids, in which a pair of eddies in the wake becomes
unstable and grow into a vortex street in the downstream region.
In contrast, we find from Fig.~\ref{f:fig2} that the alternate vortices
are directly created at the obstacle, implying that the mechanism of
vortex street formation in Fig.~\ref{f:fig2} may be different from that
in classical fluids.

\begin{figure}[t]
\includegraphics[width=8cm]{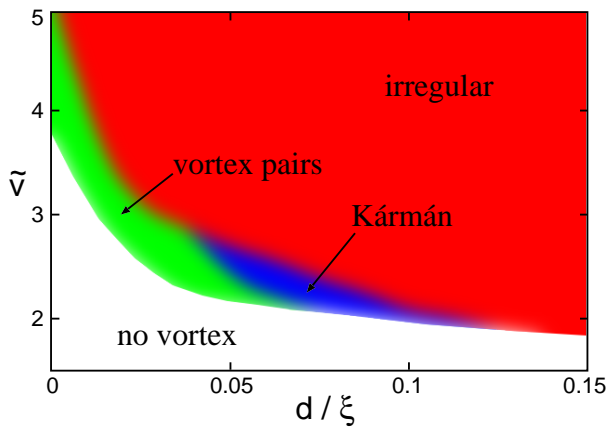}
\caption{
(Color) Dependence of the patterns of wakes on the normalized Gaussian
width  $d / \xi$ and velocity $\tilde v$ of the potential.
The green, blue, and red regions correspond to the flow patterns shown
in Figs.~\ref{f:fig1} (a), \ref{f:fig1} (b), and \ref{f:fig1} (c),
respectively.
The white region corresponds to stationary laminar flow.
}
\label{f:fig3}
\end{figure}
We systematically performed numerical simulations for various values of
$d$ and $v$ to determine the parameter regions for the types of wakes in
Fig.~\ref{f:fig1}.
Figure~\ref{f:fig3} shows a rough sketch of each parameter region.
The regions of the periodic behaviors shown in Figs.~\ref{f:fig1} (a)
and \ref{f:fig1} (b) are located between the regions of steady laminar
flow (white region in Fig.~\ref{f:fig3}) and irregular flow (red).
We note that the parameter region for vortex street formation is rather
restricted, $0.04 \lesssim d / \xi \lesssim 0.13$ and $1.9 \lesssim
\tilde v \lesssim 2.8$, where $\tilde v = v [10^3 m / (g n_0)]^{1/2} /
(2 \pi)$.
This is in contrast with classical fluids, in which the von K\'arm\'an
vortex street emerges for a wide range of Reynolds number.

\begin{figure}[t]
\includegraphics[width=8cm]{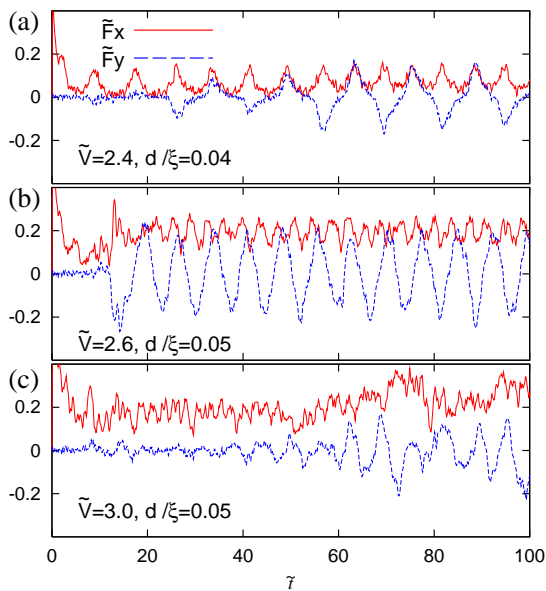}
\caption{
(Color online) Time evolution of the normalized drag force
$\tilde{\bm{F}} = \bm{F} \hbar m^{-1/2} (10^{-3} g n_0)^{-3/2}$.
The solid and dashed lines show $\tilde{F}_x$ and $\tilde{F}_y$.
The parameters used in (a)-(c) are the same as those in
Figs.~\ref{f:fig1} (a)-(c), respectively.
Time is normalized as $\tilde t = 10^{-3} t g n_0 / \hbar$.
}
\label{f:fig4}
\end{figure}
Figure~\ref{f:fig4} shows the drag force on the obstacle potential given
by $\bm{F} = \partial_t \int d\bm{r} \psi^* (i \hbar \bm{\nabla})
\psi$.
The initial state is the stationary state for $v = 0$.
At $t = 0$ the potential starts to move at a velocity $v$.
Figure~\ref{f:fig4} (a) corresponds to the vortex-antivortex pair
creation in Fig.~\ref{f:fig1} (a).
For $\tilde t \lesssim 20$, $F_x$ oscillates while $F_y \simeq 0$,
indicating that the vortex-antivortex pairs are shed from the potential
symmetrically.
For $\tilde t \gtrsim 20$, the vortex pairs begin to incline as in
Fig.~\ref{f:fig1} (a), and $F_y$ also starts to oscillate.
Figure~\ref{f:fig4} (b) corresponds to the vortex street formation,
where both $F_x$ and $F_y$ oscillate for $\tilde t \gtrsim 20$.
The oscillation in $F_y$ in Figs.~\ref{f:fig4} (a) and \ref{f:fig4} (b)
is due to the alternate shedding of vortices and hence its frequency is
half the vortex shedding frequency, i.e., the frequency of $F_x$.
It is interesting to note that $F_y$ in Fig.~\ref{f:fig4} (c)
oscillates, even though Fig.~\ref{f:fig1} (c) does not seem to have
periodicity.
A similar phenomenon is also observed in classical fluids with a large
Reynolds number~\cite{Roshko}.

\begin{figure}[t]
\includegraphics[width=8cm]{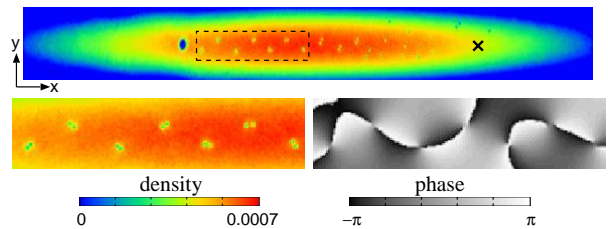}
\caption{
(Color) Density profile of a BEC of $^{87}{\rm Rb}$ atoms confined in a
 harmonic potential at $t = 330$ ms.
The Gaussian potential is initially located at the marked position
 ($\times$) and starts to move in the $-x$ direction at a constant
 velocity.
The density and phase profiles in the dashed region are magnified in the
 lower panels.
The field of view is $430 \times 43$ $\mu {\rm m}$ for the main panel
 and $86 \times 21$ $\mu {\rm m}$ for the lower panels.
The color bar is normalized by $N m \omega_y / \hbar$.
}
\label{f:fig5}
\end{figure}
Next, we study a realistic system confined in a trapping potential.
A BEC stirred by a moving obstacle potential has been studied
experimentally in Refs.~\cite{Raman,Onofrio,Neely}.
We consider a situation in which a BEC of $^{87}{\rm Rb}$ atom is
confined in a harmonic potential $m (\omega_x^2 x^2 + \omega_y^2 y^2 +
\omega_z^2 z^2) / 2$ with $(\omega_x, \omega_y, \omega_z) = 2 \pi \times
(2.5, 25, 1250)$ Hz.
The number of atoms is $N = 10^6$ and the $s$-wave scattering length is
$100 a_{\rm B}$ with $a_{\rm B}$ being the Bohr radius.
The condensate is tightly confined in the $z$ direction and the system
is effectively two dimensional.
The obstacle potential is produced by a blue-detuned Gaussian laser beam
with $d = 2.16$ $\mu {\rm m}$ and $V_0 = 100 \hbar \omega_y$, which is
initially located at $x = 130$ $\mu {\rm m}$ and $y = 0$ at $t = 0$ and
moves in the $-x$ direction at a velocity $v = 0.68$ ${\rm mm} / {\rm
s}$ for $t > 0$.
Figure~\ref{f:fig5} shows the density and phase profiles at $t = 330$
ms.
We find that a vortex street is generated behind the moving potential,
as in Fig.~\ref{f:fig1} (b), which confirms that our finding can be
experimentally observed in an inhomogeneous system.
The ratio $b / \ell$ in Fig.~\ref{f:fig5} is about $0.22$, which
deviates from 0.28 due to the finite size effect.

A future prospect of this study is to provide a mathematical description
of the symmetry breaking instability and clarify the mechanism of the
alternate creation of the vortex pairs as shown in Fig.~\ref{f:fig2}.
In classical viscous fluids, the transition from stationary flow to a
vortex street is characterized by the Hopf bifurcation and described by
the Stuart-Landau model~\cite{Provansal}.
A similar approach may be applied to the present phenomenon.
Various shapes of obstacle potentials and three dimensional dynamics
also merit further study.

In conclusion, we have shown that pairs of quantized vortices shed from
an obstacle potential moving in a BEC alternately align and survive for
a long time [Figs.~\ref{f:fig1} (b) and \ref{f:fig2}], which closely
resembles the von K\'arm\'an vortex street in classical fluids.
We have obtained the parameter region in which a vortex street emerges
(Fig.~\ref{f:fig3}) and calculated the drag force (Fig.~\ref{f:fig4}).
We have shown that vortex street formation can be observed in a
trapped BEC disturbed by a blue-detuned laser beam (Fig.~\ref{f:fig5}).
Since the von K\'arm\'an vortex street typifies the great diversity of
classical fluid dynamics, the emergence of the von K\'arm\'an vortex
street in a BEC implies that a rich variety of phenomena is still
unrevealed in quantum hydrodynamics.

We thank T. Miyazaki and N. Takahashi for their valuable comments.
This work was supported by the Ministry of Education, Culture, Sports,
Science and Technology of Japan (Grants-in-Aid for Scientific Research,
No.\ 17071005 and No.\ 20540388).

\end{document}